\documentstyle[12pt,preprint,aps,floats,epsfig]{revtex}
\input epsf 
\begin{document}
%\twocolumn [ \hsize\textwidth\columnwidth\hsize\csname @twocolumnfalse\endcsna%me
\title{Neutrino Emission from Goldstone Modes in Dense Quark Matter}

\author{Prashanth Jaikumar$^1$, Madappa Prakash$^1$, 
        and Thomas Sch\"afer$^{1,2,3}$}
\address{$^1$Department of Physics \& Astronomy, SUNY at Stony Brook, 
         NY 11794\\ 
         $^2$Department of Physics, Duke University, Durham, NC 27708\\
         $^3$Riken-BNL Research Center, Brookhaven National Laboratory, 
         Upton, NY 1197}
%\date{\today}
\maketitle
\begin{abstract}
We calculate neutrino emissivities from the decay and scattering of
Goldstone bosons in the color-flavor-locked (CFL) phase of
quarks at high baryon density.  Interactions in the CFL phase are
described by an effective low-energy theory.
For temperatures in the tens of keV range, relevant to the long-term 
cooling of neutron stars, the emissivities involving Goldstone bosons
dominate over those involving quarks, because gaps in the CFL phase 
are $\sim 100$ MeV while the masses of Goldstone modes are on the order 
of 10 MeV. For the same reason, the specific heat of the CFL phase 
is also dominated by the Goldstone modes.  Notwithstanding this, 
both the emissivity and the specific heat from the massive modes
remain rather small, because of their extremely small number
densities.  The values of the emissivity and the specific 
heat imply that the timescale for the cooling of the CFL core 
is $\sim 10^{26}$~y, which makes the CFL phase invisible as the
exterior layers of normal matter surrounding the core will continue 
to cool through significantly more rapid processes. If the CFL phase
appears during the evolution of a proto-neutron star, neutrino
interactions with Goldstone bosons are expected to be significantly 
more important since temperatures are high enough ($\sim
20-40$ MeV) to admit large number densities of Goldstone modes.


\end{abstract}

%------------------------------END OF TITLE AND ABSTRACT--------------
%\newpage
%\tableofcontents
\newpage
%-----------------------------BEGIN INTRODUCTION-------------------

\section{Introduction}
\label{sec_intro}
Neutrino emission and interactions in matter at high baryon
density play crucial roles in astrophysical phenomena such as core
collapse supernovae and neutron stars \cite{PLSV01}.  Neutrinos drive
supernova dynamics from beginning to end: they become trapped within
the star's core early in the collapse, forming a vast energetic
reservoir, and their eventual emission from the proto-neutron star is
prodigious enough---containing nearly all the energy ($\sim 10^{53}$
ergs) released in the explosion---to dramatically control subsequent
events \cite{Bur00}.  The $\nu$-luminosity and the time scale over
which $\nu$s remain observable from a proto-neutron star are also
governed by charged and neutral current interactions involving matter
at high baryon density \cite{PNSs}.  The long-term cooling of a
neutron star, up to a million years of age, is controlled by
$\nu$-emissivity  and the specific heat of the densest parts of
the star; during this time the star is observable through photon emissions,
which may allow us to determine the star's mass, radius, and internal
constitution \cite{cool}.  The tabulation of surface temperatures and
ages of neutron stars is currently one of the primary goals of X-ray
astronomy.

 	Collins and Perry \cite{CP75} noted that the superdense matter
in neutron star cores might consist of weakly interacting quarks
rather than of hadrons, due to the asymptotic freedom of
QCD. Asymptotic freedom implies that at very high baryon density, for
which the baryon chemical potential $\mu_B\gg\Lambda_{QCD}$, QCD is
amenable to perturbative techniques. However, recent studies have
shown that the naive ground state of the system, a Fermi liquid of
weakly interacting quarks, is unstable with respect to the formation
of diquark condensates \cite{Frau_78,Barrois:1977xd,Bar_79,Bailin:1984bm}.  
Attractive
interactions induce the quarks to pair which results in a gap in their
excitation spectrum that may be as large as $\Delta\simeq 100$ MeV
\cite{Alford:1998zt,Rapp:1998zu}.  The quark Cooper pairs are
anti-symmetric in color. If only two flavors participate, the pair
wave functions have the structure $\epsilon^{abc}u^b d^c$, where
$a,b,~\rm{and}~c$ are color indices. This implies that color $SU(3)$
is broken to $SU(2)$ by a Higgs mechanism and that one type of quark
and three types of gluons remain massless. These degrees of freedom
will dominate the low energy excitations of the system\footnote{ If
the temperature is very small, $T<1$ keV, the ungapped quark may
acquire a gap and the gluons will be confined into glueballs.}. If the
baryon chemical potential $\mu_B$ is very large, we expect that all
three light flavors participate.  In this case, the wave function of
the Cooper pairs is expected to be color-flavor locked
\cite{alf,rapp,Schafer:1999fe,Evans:1999at}.  Color-flavor locking
implies that all quarks and gluons acquire a gap.

   	Because the critical temperature associated with color
superconductivity is large, we expect that quark matter existing in
the core of a neutron star  will be in a superconducting
state. This leads to the question whether one can infer the presence
of a color superconducting core from astronomical observation, and
whether such observations can distinguish among the different color
superconducting phases.

  	Color superconductivity is a Fermi surface phenomenon and its
effect on the equation of state is expected to be small.  The impact
of a diquark condensed phase on the long-term cooling of neutron stars
was studied in~\cite{Blaschke,Prakash}.  These works explored the
consequences of neutrino emission processes that involved gapped
quarks.  The cooling of a pure quark star, considered in
\cite{Blaschke}, implied that cooling would occur very fast, indeed
too fast to be consistent with the existing X-ray data. In
\cite{Prakash}, in which a hybrid star with a mixed phase of hadrons
and quarks was considered, large quark gaps rendered quark matter
invisible and vanishing quark gaps led to cooling behaviors which were
indistinguishable from those of normal stars.  A possible connection
between the occurrence of quark matter in neutron stars and neutron
star glitches was proposed in~\cite{Alford}.  Neutrino diffusion in
the two-flavor superconducting (2SC) condensed phase was investigated
in ~\cite{carter}.

In this paper, we focus our attention on the long-term cooling stage,
which begins when the star's temperature has dropped to a few tenths
of an MeV ($\sim 10^9$ K or $T_9=1$) and when most of its lepton
content has been lost due to neutrino radiation during the so-called
proto-neutron star (PNS) evolution that lasts about a minute or so.
The extent of a quark phase is maximized when the trapped neutrinos
have left the star \cite{trap}.  Thereafter, the star continues to
lose its energy by radiating low-energy neutrinos until it enters a
photon cooling epoch and its surface temperature may be estimated by
the detection of X-rays.  Coupled with an independent estimate of the
star's age, the cooling history of the star may be inferred.

 	In a phase in which all quarks are gapped, the neutrino
emissivity from the direct Urca processes $f_1 + \ell \rightarrow f_2
+ \nu_\ell,~f_2 \rightarrow f_1 + \ell + \bar\nu_\ell$, where $f_1$
and $f_2$ are quarks and $\ell$ is either an electron or muon, are
strongly suppressed due to the energy cost ($\sim \Delta$) involved in
breaking the pair. (The modified Urca process, which contains an
additional quark in both the entrance and exit channels, is
additionally suppressed.) However, there are other mechanisms for
neutrino emission from this phase.  One such process is the pair
breaking and recombination of quarks~\cite{jai}. This process can
dominate over conventional cooling modes just below the transition
temperature $T_c \sim \Delta/2$.  We demonstrate here that as the
temperature falls below $T_c$, neutrino emission will be dominated by
decay of massless or almost massless collective modes, hereafter
denoted by $\phi$, that are associated with the breakdown of a global
symmetry.  The low energy excitations of CFL quark matter are an
exactly massless mode associated with the breakdown of the $U(1)$ of
baryon number and a very light octet and singlet of Nambu-Goldstone
bosons associated with the breaking of the exact $SU(3)_L\times
SU(3)_R$ chiral and approximate $U(1)_A$ axial symmetry.

 	The purpose of this work is to provide benchmark estimates of
the neutrino emissivity from the electroweak decays of Goldstone
modes. In Sec. II, we employ an effective theory of CFL Goldstone
modes coupled to electroweak gauge fields.  Specifically, we calculate
the neutrino emissivity from the processes
\begin{enumerate}
\item $\pi^{\pm}\rightarrow l^{\pm} + \bar{\nu_{l}}$,   
$K^{\pm}\rightarrow l^{\pm} + \bar{\nu_{l}}$,
\item $\pi^0(\eta,\eta')\rightarrow \nu+\bar{\nu}$\,, \quad {\rm and}   
\item $\phi+\phi\to \phi +\nu+\bar{\nu}$ \,,
\end{enumerate}
and compare our results to those from previously studied
$\nu-$emission processes.  In Sec. III, we calculate the photon
emissivity from the pion electromagnetic radiative two-photon decay,
$\pi^0\rightarrow \tilde\gamma+\tilde\gamma$. This is followed by a
discussion of photon-Goldstone boson interactions in the CFL phase.
The cooling behavior of a neutron star is also governed by the
specific heat of the ambient matter.  We therefore assess the
contribution of the Goldstone modes in Sec. IV.  Our principal
findings are summarized in Sec. V.  Utilizing our results for the
emissivity and specific heat, we estimate the time scale over which a
CFL phase core cools in the concluding Sec. VI. Here, we also comment
on the expected role of neutrino interactions with Goldstone bosons in
the evolution of a PNS, in which temperatures in the range of 20-40
MeV are encountered.

\section{Electroweak Interactions of Goldstone Modes in the CFL Phase}
\label{sec-ew}

 The low energy excitations of the CFL phase are two singlet 
modes associated with $U(1)_B$ and $U(1)_A$ symmetry and
an octet of Goldstone modes associated with chiral symmetry
breaking. The octet is described by a low energy theory
that bears strong resemblance to chiral perturbation 
theory~\cite{ismail,gatto,Son:1999cm}. The Goldstone modes are
parametrized by a $3\times3$ unitary matrix $U$ which is a 
color singlet, transforming under $SU(3)_L\times SU(3)_R$ 
as $U\rightarrow g_LUg_R^{\dag}$, where $U$ is related 
to axial-like fluctuations of the left and right
handed diquark fields:
\begin{equation}
 L^{ai}\sim\epsilon^{abc}\epsilon^{ijk}\langle q_L^{bj}q_L^{ck} 
 \rangle^{*},\quad R^{ai}\sim\epsilon^{abc}\epsilon^{ijk} 
 \langle q_R^{bj}q_R^{ck}\rangle^{*},\quad {\rm and} \quad U=LR^{\dag}\quad.
\end{equation}
The fields $L$ and $R$ carry color $(ijk)$ and flavor $(abc)$, 
and transform under $g_f \subset SU(3)_f ~{\rm and}~ g_c\subset SU(3)_C$  
(where $f$ denotes left or right handed flavor) as
\begin{equation}
L\rightarrow g_LLg_C^{\dag}\,\quad {\rm and} 
\quad R\rightarrow g_RRg_C^{\dag}\,,
\end{equation}
respectively. The low energy effective theory is governed 
by the Lagrangian
\begin{equation}
{\cal L}_{eff}=\frac{f_{\pi}^2}{4}{\rm Tr}\biggl[\partial_0
 U\partial_0 U^{\dag} - v_{\pi}^2\partial_i U \partial_i
 U^{\dag}\biggr] +\ldots \, ,
 \label{leff}
\end{equation}
where $v_{\pi}$ is the Goldstone boson velocity and we have
suppressed mass terms and higher derivative terms. The chiral 
field $U$ can be related to the octet meson field by $U=\exp
(i\pi^a\lambda^a/f_\pi)$, where the $SU(3)$ generators $\lambda^a$ 
are normalized as ${\rm tr}[\lambda^a \lambda^b]=2\delta^{ab}$.

 In the weak coupling limit, we have $v_\pi^2=1/3$. As a consequence 
of the breaking of Lorentz invariance in matter, there are two pion 
decay constants, $f_T$ and $f_S$, which characterize the coupling of 
the pion to the temporal and the spatial components of the 
axial-vector current. The effective Lagrangian in Eq.~(\ref{leff})
implies that $f_T=f_\pi$ and $f_S=v_{\pi}^2f_\pi$. This result 
is consistent with axial-vector current conservation in the limit
$m_\pi\to 0$. We observe that $f_T\omega^2-f_S{\bf q}^2=0$ 
for an on-shell pion. 

 The extension to include electroweak interactions is achieved by
minimal coupling~\cite{casal}:
\begin{eqnarray}
D_{\mu}L &=& \partial_{\mu}L-ieA_{\mu}QL-\frac{ig}{\sqrt{2}} 
  \biggl[W_{\mu}^{+}\sigma^{+}+W_{\mu}^{-}\sigma^{-}\biggr]L
  -\frac{ig}{2{\rm cos}\,\theta_W}Z_{\mu}^0\biggl[\sigma^{3}-2Q{\rm
   sin}^{2}\theta_{W}\biggr]L \,, \label{ewgauge} \\ \nonumber
D_{\mu}R &=& \partial_{\mu}R-ieA_{\mu}QR +\frac{ig}{2{\rm cos}\,
  \theta_W}Z_{\mu}^0\biggl[2Q{\rm sin}^{2}\theta_{W}\biggr]R \,.
\end{eqnarray}
Here, $A$ is the electromagnetic field\footnote{The $A$ field may be
re-expressed in terms of the physical  (massive) gluon  and physical
 (massless) photon  that arise as a consequence of photon-gluon mixing
in the CFL phase.}, which couples  with strength $Q={\rm
diag}\,(2/3,-1/3,-1/3)$ to the $u,d,$ and $s$ quarks, $W$ and $Z$ 
are the $SU(2)$ electroweak gauge fields, $g$ is the weak coupling 
constant, and $\theta_W$ is the Weinberg angle. The $\sigma$ 
matrices are the Pauli matrices of $SU(2)$ embedded 
into flavor $SU(3)$: 
\begin{equation}
\sigma^{+}=\bordermatrix{
&  &   & \cr
&0 & 1 & 0 \cr
&0 & 0 & 0 \cr
&0 & 0 & 0 \cr}
,\quad \sigma^{-}=\bordermatrix{
&  &   & \cr
&0 & 0 & 0 \cr
&1 & 0 & 0 \cr
&0 & 0 & 0 \cr}
,\quad {\rm and} \quad
\sigma^3=\frac{1}{2}\bordermatrix{
&  &   & \cr
&1 & 0 & 0 \cr
&0 & -1 & 0 \cr
&0 & 0 & -1 \cr} \,. \nonumber
\end{equation} 

	In the charged current coupling, we have only taken into 
account the first two flavors~\cite{casal}. We will return later to comment 
on charged current decays of kaons. We observe that 
the CFL pions, kaons, and etas have charged and neutral 
current interactions that are very similar to 
those of ordinary mesons. 
An important difference, however, is the fact that CFL mesons are very
light, $m_{GB} \simeq 10$ MeV \cite{Son:1999cm}; see
\cite{Schafer:2002ty,thomas} for a recent discussion. This
implies that the pion and kaon are lighter than the muon and
dominantly decay into electrons and neutrinos. Another difference, as
compared to the zero density case, is the fact that the Goldstone boson
dispersion relation is modified from its vacuum form:  
\begin{equation}
\label{disp}
E_p= \mu_{eff}+\sqrt{m_{GB}^2+v_\pi^2 p^2}\,.
\end{equation}
Here, $\mu_{eff}$ is an effective chemical potential
that depends on the quark masses \cite{Bedaque:2001je}.
For pions, $\mu_{eff}\simeq 0$, but for kaons $\mu_{eff}$
is sizeable and may have consequences. 
The modified dispersion relation 
Eq.~(\ref{disp}) leads to a number of unusual 
effects. One of them, the fact that a very fast 
pion becomes absolutely stable, was already discussed
in \cite{Zarembo:2000pj}. Here, we will encounter
a second one, the presence of the ``helicity forbidden''
decay $\pi^0\to\nu\bar{\nu}$.

\subsection{Neutrino Emissivity from $\pi^{\pm}(p)\rightarrow 
l^{\pm}(p_1) + \bar{\nu_{l}}(p_2)$ } 
\label{sec_pip}

  	Substituting Eq.~(\ref{ewgauge}) into the effective Lagrangian, 
we obtain the relevant interaction term for charged pion
decay:
\begin{equation}
 {\cal L}_{W\pi}=\frac{gf_T}{2}\biggl(W_{0}^{-}\partial^{0}\pi^{+} + 
     W_{0}^{+}\partial^{0}\pi^{-}\biggr)
     - \biggl( f_T\rightarrow f_S, 0\rightarrow i\biggr)\quad.
\end{equation}
The pion decay constant $f_{\pi}$ has been estimated using various
methods in the literature~\cite{Son:1999cm,Zd}. In
Ref.~\cite{Son:1999cm}, it was computed by matching a  
chiral effective theory with a microscopic theory of quasiparticles 
and holes near the Fermi surface. Explicitly,
\begin{equation}
 f_{\pi}^2=\frac{21-8\,{\rm ln}\,2}{18}~\frac{\mu^2}{2\pi^2}\quad.
\end{equation}
In the following, we will outline the calculation of the $\nu-$emissivity 
for the $\pi^{-}$ decay. The $\pi^{+}$ decay gives exactly the same
result.
The $T$-matrix element for the weak decay process is 
\begin{equation}
 \langle l^{-}(p_1)\bar{\nu}_l(p_2)|T|\pi^{-}(p)\rangle = 
    (2\pi)^4\delta^4(p-p_1-p_2)G_F
       \bar{u}_l(p_1)(\gamma^{0}p_0 f_T-{\bf\gamma\cdot p}f_S)
         (1-\gamma_5)v_{\nu_{l}}(p_2)\quad,
\end{equation}
where we have made the identification $G_F/\sqrt{2}=g^2/(8M_W^2)$. Using the
equation of motion for the leptonic fields, the neutrino emissivity
from this process is given by 
\begin{eqnarray}
\label{emisspi} 
 \epsilon_{\pi} &=&\int\frac{d^3p}{(2\pi)^32\omega_p}n_B(\omega_p)\int
  \frac{d^3p_1}{(2\pi)^32\omega_{p_1}}\tilde{n}_F(p_1) \int 
  \frac{d^3p_2}{(2\pi)^32\omega_{p_2}}(\omega_{p_2})(2\pi)^4\delta^4(p-p_1-p_2)
          \sum_{spin}|M|^2, \\
\label{M}
 & &\sum_{spin}|M|^2 = 8G_F^2\biggl(|\xi|^2 p_1 \cdot p_2 + 
          |\chi|^2 \left\{\omega_p^2(\omega_{p_1}\omega_{p_2}+
                      {\bf p}_1 \cdot {\bf p}_2)
          +m_e^2(p_1\cdot p_2-2\omega_p\omega_{p_2})\right\} \nonumber \\
 & &  \hspace{3cm}\mbox{}
         + 2|\xi||\chi|m_e(\omega_p\omega_{p_2}-p_1 \cdot p_2)\biggr)\,, 
\end{eqnarray}
where $\omega_p=\sqrt{m_{GB}^2+v_\pi^2 p^2}$, $\xi = -im_ef_T$ and $\chi=-i(f_T-f_S)$. Here, $m_e$ is the 
electron mass, $n_B(\omega_p)$ denotes the Bose
occupation factor for the pion, and $\tilde{n}_F(p_1)=1-n_F(p_1)$ 
is the Pauli blocking factor for the outgoing electron.  For the temperatures 
of relevance to long-term cooling,  
Pauli blocking of the outgoing neutrino may be neglected. 
For bulk CFL matter at $T=0$, the electron chemical 
$\mu_e \rightarrow 0$ \cite{Bedaque:2001je,Rajagopal:2000ff}. At non-zero
temperature, there is a small electron chemical potential due 
to the fact that the energies of positively and negatively
charged Goldstone modes are not the same. For $T<1$ MeV, we 
find $\mu_e\ll T$ and the effect on neutrino emission is 
negligible. In proto-neutron stars, 
CFL phases with charged meson condensates can also exist
\cite{Kaplan:2001qk}. Since our focus here is on 
long-term cooling, we do not consider these possibilities.
Finally, we expect that CFL matter inside a neutron star
is in contact with a hadronic or quark phase that has a 
large electron chemical potential. As discussed in 
\cite{Alford:2001zr}, this will lead to a thin charged surface
layer which shields the CFL phase. In the following, we
will assume that $\mu_e=0$ in regions well separated from this layer.\\

  The squared matrix element in Eq.~(\ref{M}) consists of three terms. 
We will evaluate the emissivities from each in turn. The first term
involves $|\xi|^2$ and has the Lorentz invariant structure 
familiar from its vacuum counterpart. This is the only
term that survives in the limit $f_T=f_S$. We also observe
that the emissivity is proportional to the electron mass squared, 
as expected. The momentum dependence is of the form 
$p_1\cdot p_2 = \omega_{p_1}\omega_{p_2}(1-v_1v_2{\rm cos}\,\theta)$ 
with $v_1=|{\bf p_1}|/\omega_{p_1},\, v_2=|{\bf p_2}|/\omega_{p_2}$, 
and $\theta=\angle ({\bf p}_1,{\bf p}_2)$. The angular integral 
gives
\begin{eqnarray}
 \int d\Omega_p&d\Omega_{p_1}&d\Omega_{p_2} 
   (1-v_1v_2\cos\theta)(2\pi)^3 
   \delta^3({\bf p}-{\bf p}_1-{\bf p}_2) \\ \nonumber
  &=& \frac{(2\pi)^5}{pp_1p_2\omega_{p_1}\omega_{p_2}} 
     \biggl(2\omega_{p_1}\omega_{p_2}-(p^2-p_1^2-p_2^2)\biggr)
	\Theta(p+p_1-p_2) \Theta(p+p_2-p_1)\Theta(p_1+p_2-p) \,,
\label{intangle}
\end{eqnarray} 
where $p$ is now to be understood as $|{\bf p}|$
{\it etc}. The triangle inequalities imposed by momentum conservation 
restrict the range of
the $p_2$ and $p$ (or equivalently $\omega_p$) integrals as follows: 
\begin{eqnarray}
\frac{\omega_p-p}{2}-\frac{m_e^2}{2(\omega_p+p)}&\leq&p_2\leq
\frac{\omega_p+p}{2}-\frac{m_e^2}{2(\omega_p-p)} \label{intlimits} \,,
\nonumber \\
m_{\pi}&\leq&\omega_p\leq \sqrt{\frac{m_{\pi}^2-m_e^2}{1-v_{\pi}^2}} \,.
\end{eqnarray}
When these conditions are satisfied, the $p_1$ integration can be done
trivially using the energy-delta function. 
The expression for the emissivity then becomes
\begin{equation}
\epsilon_{\xi} = \frac{A}{64{\pi}^3v_{\pi}^2}
         \int_{m_{\pi}}^{\omega_p^{max}}d\omega_p\, n_B(\omega_p)
	 \biggl(\frac{m_{\pi}^2}{v_{\pi}^2}-m_e^2
           +\frac{(v_{\pi}^2-1)}{v_{\pi}^2}\omega_p^2\biggr)        
 	\int_{p_2^{min}}^{p_2^{max}}dp_2\, p_2 
         \frac{1}{1+\exp{(p_2-\omega_p)/T}}
         \,, \label{epspi}
\end{equation}
where $A=8G_F^2f_T^2\,m_e^2$ and the limits 
$\omega_p^{max},~p_2^{min},~{\rm and}~p_2^{max}$ are defined
by~Eq.~(\ref{intlimits}). In obtaining~Eq.~(\ref{epspi}), we have
utilized the pion dispersion relation 
$\omega_p^2=v_{\pi}^2p^2+m_{\pi}^2$ to change the
variable of integration as $pdp/\omega_p=d\omega_p/v_{\pi}^2$. 
We have also used the identity $\tilde{n}_F(x)
=n_F(-x)$.
For long-term cooling, temperatures range from tens of keV to eV, much
less than either the pion mass or electron mass ($\sim 10$ MeV and 0.51
MeV, respectively). In the following, we will therefore 
calculate the emissivity in the low temperature limit when $m_\pi/T$ and 
$m_e/T\gg 1$. Unless the pion mass becomes very small, $m_\pi
\simeq m_e$, we can also neglect the electron mass in the 
integrand of Eq.~(\ref{epspi}). 
 Within the range
of the $\omega_p$ integration, we always have $p_2\leq \omega_p$. In 
the low temperature limit, $m_\pi/T\gg 1$, we can replace the
Pauli-blocking factor $n_F(-x)$ by unity. In this limit, we can also
replace the Bose distribution function for the pion by the Boltzmann
distribution. Introducing the scaled variables
$x=\omega_p/T, ~y=p_2/T,~{\rm and}~ \psi=m_{\pi}/T$,
Eq.~(\ref{epspi}) now reads
\begin{equation}
\epsilon_{\xi} = \frac{AT^5}{64{\pi}^3v_{\pi}^2}\int
_{\psi}^{\psi_{max}}dx~e^{-x}\biggl(\frac{\psi^2}{v_{\pi}^2} -
\left(\frac{1-v_{\pi}^2}{v_{\pi}^2}\right)x^2\biggr)
\int_{y_{min}}^{y_{max}}dy~y \,,
\end{equation}
where $\psi_{\max}=\psi/\sqrt{1-v_{\pi}^2}$. The $y$ integration
yields $x\sqrt{x^2-\psi^2}/2v_{\pi}$ with the result that
\begin{equation}
\epsilon_{\xi} = \frac{AT^5}{64{\pi}^3v_{\pi}^5}\int
_{\psi}^{\psi_{max}}dx~
 e^{-x}~\frac{x\sqrt{x^2-\psi^2}}{2}\biggl(\psi^2-(1-v_{\pi}^2)x^2\biggr) 
\quad.\label{intpsi}
\end{equation}
As $\psi$ is parametrically large, most of the contribution to the
integral comes near the lower limit. The upper limit can be extended
to infinity with little impact on the numerical value of the
integral. Using the substitution $x=\psi~{\rm cosh}~\theta$ (the limits
on $\theta$ now run from 0 to $\infty$), the definition of the
modified Bessel function of the second kind $K_\nu$ and their
recursion relation, we finally obtain
\begin{equation}
\epsilon_{\xi} =
\frac{A}{64{\pi}}m_{\pi}^2
\biggl[\frac{1}{v_{\pi}^3}\biggl(\frac{m_{\pi}T}{2\pi}\biggr)^{3/2}
 e^{-m_{\pi}/T}\biggr] 
 = \frac{A}{64{\pi}}m_{\pi}^2~n_\pi
\label{piemiss1} \,,
\end{equation}
where we have used the asymptotic form $K_{\nu}(z)=\sqrt{\pi/2z}~{\rm
e}^{-z}$ valid for any $\nu$ when $z\gg 1$.  The factor A is
proportional to the interaction strength and the factor in the
square bracket is the number density $n_\pi$ of thermal non-degenerate
pions. Note that the dispersion relation in Eq.~(\ref{disp}) introduces an 
extra factor of $1/v_\pi^3$ in the number density of pions relative to the 
case with $v_\pi=1$. 

The other two contributions to the emissivity in Eq.~(\ref{emisspi})
are evaluated along similar lines. The result for the mixed term
($|\xi||\chi|$) and the quadratic term ($|\chi|^2$) are
\begin{eqnarray}
\epsilon_{\xi\chi} = 
\frac{B}{64{\pi}}m_{\pi}^2~n_\pi \quad {\rm and} \quad  
%\label{piemiss2}\\
\epsilon_{\chi} =
\frac{C}{64{\pi}v_\pi^2}m_{\pi}^2~n_\pi \,, 
 \label{piemiss3}
\end{eqnarray}
 respectively, with $B=16G_F^2f_T(f_T-f_S)m_e^2$ and
$C=16G_F^2(f_T-f_S)^2m_{\pi}T$. We observe that the coefficient $C$ is 
independent
of the electron mass. The term involving $C$ arises purely as a
consequence of Lorentz symmetry breaking in dense matter. 
We also note that all emissivities appear to diverge as
$v_\pi\to 0$, which is the case near a phase transition. 
This is an artifact of the approximation $\psi_{max}
\to\infty$ in Eq.~(\ref{intpsi}). At fixed $T$, all rates are
finite as $v_\pi\to 0$. 
Combining Eqs.~(\ref{piemiss1}) and (\ref{piemiss3}), 
and using the perturbative result $v_\pi^2=1/3$, we obtain the 
neutrino emissivity from the electroweak decay of CFL pions  as
\begin{eqnarray}
  \epsilon_\pi &=& \epsilon_\xi + \epsilon_{\xi\chi} + \epsilon_\chi 
\nonumber \\  
	&=& 	\frac{1}{8\pi}\,
    (G_F^2\, f_{\pi}^2\, m_e^2)\, m_\pi^2\, n_{\pi}\,
    \left( 1 + 2\left(\frac{\delta f_\pi}{f_\pi}\right)
      + \frac{2m_\pi T}{v_{\pi}^2m_e^2}\left(\frac{\delta f_\pi}{f_\pi}\right)^2
   \right)\,,
\label{emisspiappx}
\end{eqnarray}
where $\delta f_\pi=f_T-f_S$. The rate from $\pi^+$ decay is 
exactly the same. This result is valid when temperatures fall 
below an MeV. Numerically, the emissivity may be expressed as
\begin{equation}
\label{eps_pipm}
\epsilon_{\pi}=1.18\times
     10^{28}\mu_{100}^2~m_{10}^2~(m_{10}T_9)^{3/2} 
     e^{-\frac{116m_{10}}{T_9}}
\biggl[1+1.14\biggl(\frac{m_\pi}{m_e}\biggr)
\biggl(\frac{T}{m_e}\biggr)\biggr] ~{\rm erg~cm^{-3}~s^{-1}} \,,
\end{equation}
where $\mu_{100}$ is the quark chemical potential in units of 100 MeV,
$m_{10}$ is the pion mass in units of 10 MeV, and $T_9=T/10^9\,
{\rm K}$.

It is interesting to compare the neutrino emissivity from pion decay
with the emissivity from the quark direct Urca process in a gapped
superfluid, as well as the quark pair breaking and formation (PBF)
process. From \cite{Iwamoto:eb,Iwa}, the emissivity from the quark
direct Urca process is given by
\begin{equation}
 \epsilon_{q\beta} \simeq 8.8\times 10^{26}\alpha_s 
  \biggl(\frac{n_B}{n_0}\biggr)
  Y_e^{1/3} T_9^6~e^{-\Delta/k_BT}~
  {\rm erg~cm^{-3}~s^{-1}} \,,
\end{equation}
where $\alpha_s=g^2/4\pi$ is the strong coupling constant, $n_B$ is the 
baryon density, $n_0=0.16~{\rm fm^{-3}}$ is the nuclear equilibrium density, 
and $Y_e=n_B/n_0$ is the electron concentration. For $n_B=5n_0$,
$\alpha_s\simeq 1$, $Y_e\sim 10^{-6}$, $\Delta\simeq 100$ MeV and 
$m_\pi\simeq 10$ MeV, we find that pion decay dominates the Urca
emissivity for $T<5\times 10^{10}$ K. The emissivity from the quark PBF, 
which is effective in rapidly cooling the star during its early 
hundreds of years, is given by~\cite{jai}
\begin{equation}
 \epsilon_q^{\nu\bar\nu} \simeq 1.4 \times 10^{20}~N_\nu T_9^7 
  F~ a_q \left(\frac{n_B}{n_0}\right)^{2/3}~
  {\rm erg~cm^{-3}~s^{-1}} \,, 
\label{epsq}
\end{equation}
where $N_{\nu}$ is the number of neutrino flavors, $F$ is a
temperature dependent function of order one that vanishes exponentially
with the gap, and $a_q$ is a flavor dependent numerical factor
of order 0.1. In the vicinity of $T_c$, there is no exponential 
suppression, but as $T\ll T_c$, the function $F$ approaches
$\exp(-2\Delta/k_BT$) so that the emissivity from PBF becomes
very small.

%----------------------\pi^0\rightarrow\nu+\bar{\nu}-----------------------

\subsection{Neutrino Emissivity from $\pi^0(\eta,\eta')\rightarrow 
\nu+\bar{\nu}$}
\label{sec_pi0}

 	In the previous section, we noticed that the emissivity for
$\pi^-\to e^-+\nu$ process in matter contains a term that is not
proportional to $m_e^2$. This implies that the neutral current decays
$\pi^0 \rightarrow \nu+\bar{\nu}$ and $\eta(\eta')\rightarrow
\nu+\bar{\nu}$, which are forbidden by helicity selection rules in
vacuum, can occur in matter. The reason these decays are forbidden in
vacuum is that the two neutrinos have opposite chirality.  Thus, the
two neutrinos have total angular momentum one in the rest frame of the
decaying meson, Hence, a scalar meson cannot decay into
$\nu\bar{\nu}$. In a dense medium, it is still true that a scalar
meson which is at rest with respect to the medium cannot decay into
$\nu\bar{\nu}$ pairs. However, boost invariance is lost.  This implies
that the wave function of a scalar meson which is moving with respect
to the medium can have higher angular momentum admixtures when viewed
in its rest frame.

 	In the following, we calculate the emissivity due
to the process $\pi^0\to\nu+\bar{\nu}$. The interaction term 
relevant for this decay channel is given by
\begin{equation}
 {\cal L}_{Z^0\pi^0} = \frac{f_T~g}{2~\cos\theta_W}
  \biggl(Z^0_0\partial^0\pi^0\biggr)
 -\biggr(f_T\rightarrow f_S, 0\rightarrow i\biggr) \,.
\end{equation}
The corresponding interaction for the $\eta$ decay has an extra 
factor of $1/\sqrt{3}$ ($1/\sqrt{6}$ for $\eta^{\prime}$ decay) 
from the normalization of the $SU(3)$ 
generators. For simplicity, we ignore $\pi^0-\eta-\eta'$ 
mixing \cite{Son:1999cm,thomas}. We will outline the emissivity 
calculation for the $\pi^0$ decay. The corresponding emissivity for 
the $\eta$ is obtained by changing the overall factor and the 
mass. The neutrino pair emissivity from $\pi^0$ decay is given
by
\begin{eqnarray}
 \epsilon_{\nu\bar{\nu}} &=& \int\frac{d^3p}{(2\pi)^32\omega_p} n_B(\omega_p)
      \int \frac{d^3p_1}{(2\pi)^32\omega_{p_1}} 
      \int \frac{d^3p_2}{(2\pi)^32\omega_{p_2}}
        (\omega_{p_1} + \omega_{p_2})
        (2\pi)^4\delta^4(p-p_1-p_2)\sum_{spin} |M|^2 
\label{emissnu}\\ 
   & & \sum_{spin}|M|^2 = 4N_{\nu}G_F^2|\chi|^2
      \omega_p^2(\omega_{p_1}\omega_{p_2}
         +{\bf p}_1\cdot{\bf p}_2) \,, \nonumber 
\end{eqnarray}
where $N_{\nu}$ counts the number of neutrino flavors. The 
emissivity is proportional to $f_T-f_S$, so it is directly
proportional to Lorentz symmetry breaking in matter. The integrals are
performed in the limit $m_e\to 0$ and $m_{\pi}/T\gg 1$ as before, and
the emissivity from this process is found to be
\begin{equation}
 \epsilon_{\nu\bar{\nu}}=\frac{N_{\nu}
	}{4\pi~v_{\pi}^2}(G_F^2(\delta f_{\pi})^2m_{\pi}^2)~Tm_{\pi}~n_{\pi}. 
\label{emissnuappx} 
\end{equation}
Numerically, the emissivity  (for one neutrino flavor) is
\begin{equation}
  \epsilon_{\nu\bar{\nu}}= 4.52\times
     10^{28}\mu_{100}^2~m_{10}^2~(m_{10}T_9)^{5/2}
     e^{-\frac{116m_{10}}{T_9}}~ {\rm erg~cm^{-3}~s^{-1}} \,.
\end{equation}

\subsection{Emissivities from Kaon and Massless Goldstone Boson Decays}
\label{sec_oth}

 In this section, we would like to provide a brief survey of 
other processes that contribute to the emissivity of CFL matter. 
Neutrino pair emission from plasmon decays in a stellar plasma, 
first proposed in~\cite{adams} and subsequently investigated in 
detail in~\cite{Bra,segel}, can be the dominant energy loss mechanism 
for very hot and dense cores of red giant stars or white dwarfs. 
Blaschke {\it et al.} \cite{Blaschke} studied the analogous plasmon 
decay of the massive gluon in the CFL phase. They found that the rate 
is suppressed by $\exp(-m_g/T)$, where $m_g \sim g\mu$ is the 
effective gluon mass. This implies that the plasmon rate is 
even more strongly suppressed than the quark Urca rate. 

 	The physical photon $\tilde{A}$ in the CFL phase is a 
linear combination of the ordinary photon ($A$) and gluon
($G$) fields
\begin{equation}
 \tilde{A}_{\mu} = A_{\mu} \cos\theta_{CFL} 
 - G_{\mu}^{8}\sin\theta_{CFL}, \hspace{1cm}
  \tan\theta_{CFL} = 2e/\sqrt{3}g \,,
\end{equation}
where $e$ and $g$ are the electromagnetic and strong coupling
constants. The massive gluon $\tilde{G}^8$ is the orthogonal
combination. In an electron plasma, the photon is dressed by 
particle-hole excitations and acquires an effective mass. As
a consequence, it can decay into $\nu\bar{\nu}$ pairs. In CFL
matter at $T=0$, the photon is dressed by both particle-hole and 
particle-particle excitations, and the dielectric constant
is larger than one, but it does not acquire a mass ~\cite{Litim,Jai}. 
As a result, it cannot decay into $\nu\bar{\nu}$ pairs. At
$T\neq 0$, the photons acquire a small mass, mainly because
of thermal $e^+e^-$ pairs, but the corresponding $\tilde\gamma
\to\nu+\bar{\nu}$ rate is very small. 

  In CFL matter, the usual spectrum of meson masses is 
partially inverted \cite{Son:1999cm,Bedaque:2001je}. In particular,
we expect that Goldstone bosons with positive strangeness, the $K^0$
and the $K^+$, are lighter than non-strange Goldstone bosons such as
pions and etas. The $K^+$ contributes to neutrino emission via the
decay $K^+\to e^+ +\bar{\nu}$. The emissivity for this process can be
computed along the same lines as the emissivity from charged
pion decay. However, there are two differences. First, kaon decay
is suppressed by the Cabbibo angle $\sin^2(\theta_C)$, where
$\theta_C\sim 15^{\circ}$. The second difference is related to
the fact that the kaon dispersion relation is modified by the
effective chemical potential term, see Eq.~(\ref{disp}). We find
\begin{equation}
 \epsilon_{K^{\pm}} =
   \frac{A_K}{64\pi} E_{K^\pm}^2n_{K^\pm}
   \left(  1 + 2\left(\frac{\delta f_\pi}{f_\pi}\pm\frac{\mu_{eff}}{E_{K^\pm}}\right)
      + \frac{2m_{K^\pm} T}{v_K^2 m_e^2}
          \left(\frac{\delta f_\pi}{f_\pi}\pm\frac{\mu_{eff}}{E_{K^\pm}}\right)^2 \right)\,, \label{Kemiss}
\end{equation}
where $A_K=8\sin^2(\theta_C)G_F^2f_T^2m_e^2$ is the effective
coupling, and $E_{K^\pm}=m_K\mp \mu_{eff}$ with $\mu_{eff}=
m_s^2/(2\mu_q)$ is the energy of a kaon at rest. Note that in the
limit $m_e\rightarrow 0$, $\delta f_\pi\rightarrow 0$, and
$v_K\rightarrow 1$, the emissivity remains finite because the terms containing
$\mu_{eff}$ breaks Lorentz invariance.  In this case, 
\begin{equation}
\epsilon_{K^{\pm}} =
   \frac{\sin^2(\theta_C)G_F^2f_T^2}{4\pi v_K^2}m_{K^\pm}n_{K^\pm}T\mu_{eff}^2
\end{equation}
For $T\sim 10^9K$, this contribution to the emissivity would dominate over that of the first two terms in Eq. (\ref{Kemiss}) that are proportional to $m_e^2$ . We note that the $K^+$ emissivity is not equal to the $K^-$
emissivity.  This implies that, strictly speaking, it is not
legitimate to ignore the effects of a non-zero electron chemical
potential.  We also note that, because of the smaller energy of a kaon
compared to a pion, kaons will likely dominate the emissivity despite
the Cabbibo suppression. Numerically, we find
\begin{eqnarray}
\epsilon_{K}=3.38\times 10^{26}\mu_{100}^2~E_{10}^2~
   (&m_{10}&T_9)^{3/2}e^{-\frac{116 E_{10}}{T_9}} \nonumber \\
   &\times&\biggl[1 + 2\left(\frac{\delta f_\pi}{f_\pi}\pm\frac{\mu_{eff}}{E_{K^\pm}}\right)
      + \frac{2m_{K^\pm} T}{v_K^2 m_e^2}
          \left(\frac{\delta f_\pi}{f_\pi}\pm\frac{\mu_{eff}}{E_{K^\pm}}\right)^2\biggr]~{\rm erg~cm^{-3}~s^{-1}} \,,
\end{eqnarray}
where $m_{10}$ and $E_{10}$ are the kaon mass and energy in
units of 10 MeV. In Sec. \ref{sec_pi0}, we saw that in CFL
matter the decay $\pi^0\to \nu+\bar{\nu}$ is allowed if the
$\pi^0$ has non-zero momentum. The analogous decay $K^0\to
\nu+\bar{\nu}$, however, remains strongly suppressed
because it requires a second order weak $s\to d$ transition.

 	Since the CFL phase is characterized by unusually light
Goldstone bosons, there is a possibility for Bose condensation to
occur. Indeed, it has been argued that CFL matter at
densities that can be achieved in neutron stars is likely to support a
$K^0$ condensate
\cite{Bedaque:2001je,Schafer:2000ew,Bedaque:2001at}. Furthermore,
because of the presence of trapped neutrinos, CFL matter in a
proto-neutron star may have charged pion or kaon condensates
\cite{Kaplan:2001qk}. In neutron matter, pion condensation (if it
occurs) leads to a substantial increase of the neutrino emissivity
through processes like $n+ \pi^+ \to n +e^++\bar{\nu}$
\cite{Bahcall:1965,Friman:1979}. In CFL matter, there are no ungapped
fermions and the analogous process is suppressed by
$\exp(-\Delta/(k_BT))$.

 	If charged pions or kaons become lighter, then the emissivity
from the decay process $\pi^\pm \to e^\pm +\bar{\nu}$ will initially be
enhanced. There is, however, no neutrino emission from the decay of a
massless charged pion or kaon in a Bose condensed phase. The reason is
that the decay of a massless boson with dispersion relation $\omega
=\frac{1}{3}|\vec{k}|$ into pairs of leptons is kinematically
forbidden. Indeed, we observe that the emissivity for the process
$\pi^\pm \to e^\pm+\bar{\nu}$ is of the form $m_\pi^{7/2}
\exp(-m_\pi/(k_BT))$ and vanishes as $m_\pi\to 0$.

 	Instead, we have to consider neutrino Bremsstrahlung in
Goldstone boson scattering, or Goldstone boson scattering followed by
the decay of a Goldstone boson into $\nu\bar{\nu}$ or $e^\pm \bar{\nu}$. In
the case of massive Goldstone bosons, these processes are suppressed by
an additional power of $\exp(-m_{GB} /(k_BT))$ with respect to the
direct Goldstone boson decay process. For massless Goldstone bosons,
however, neutrino emission is dominated by Goldstone boson scattering.

 	In CFL matter, there is always at least one exactly massless
Goldstone boson which is associated with the breaking of the $U(1)_B$
of baryon number. The $U(1)_B$ Goldstone boson contributes to the
neutrino emissivity through the process $\phi+\phi\to \phi
+\nu+\bar{\nu}$. The coupling of $\phi$ to the $Z$-boson is given by
$(g/(12\cos(\theta_W)) Z_\mu f\partial^\mu \phi$ where $f$ is the $U(1)_B$
decay constant \cite{Son:1999cm}.  The $\pi\pi$ and $KK$ scattering
amplitudes are fixed by the leading order $O(p^2)$ chiral
Lagrangian. The $\phi\phi$ scattering amplitude, on the other hand,
vanishes at leading order in the low energy expansion. 
The $\phi\phi$ scattering amplitude is on the
order of $A\sim p^4/(f^2\mu^2)$~\cite{Son02}. 
The emissivity will then scale as
\begin{equation}
 \epsilon_{GB} \sim   \frac{G_F^2}{f^2\mu^4} T^{15} \,.
\end{equation}
Including numerical factors associated with the scattering amplitude 
and the phase space integral, 
\begin{equation}
 \epsilon_{GB} \simeq 
	  10^{-11} \times T_9^{15} \mu_{100}^{-6}~
		 {\rm erg~cm^{-3}~s^{-1}} \,. 
\end{equation}
This estimate is adequate for $T\ll\Delta$. 
For the range of temperatures considered in this work,  
this process is not likely to play an important role. 
Since there is no exponential suppression, it will, however, 
dominate neutrino emission at very late times.

%--------------- pi^0 -> gamma gamma---------------------------------------

\section{Photon Emissivity and Propagation in the CFL phase}
\label{sec_pi0gg}

 	So far, we have identified sources for neutrino emission from
CFL quark matter. Another interesting issue is the photon emissivity
from this phase. A first step towards computing the emissivity of
photons from the CFL phase was taken in~\cite{Jai}, where the
contributions from pair-correlated quarks and  $q\bar{q}$ annihilations
were found to be comparable in magnitude to the rates from a hot
hadron gas, at temperatures of several tens of MeVs. At keV
temperatures, however, the rates are vanishingly small due to the fact
that quark gaps are of order 100 MeV. 

Another source of photons in CFL matter is the anomalous
$\pi^0\to\tilde{\gamma}+\tilde{\gamma}$ decay, which proceeds via the
non-abelian axial anomaly.  The photon emissivity due to this process 
is given by
\begin{eqnarray}
 \epsilon_{\tilde\gamma} 
  &=& \frac{1}{2}\int \frac{d^3q\,n(\omega_q)}{2\omega_q(2\pi)^3}\,
    \omega_q \int\,\frac{d^3p\,(1+n(\omega_p))}{2\omega_p(2\pi)^3}\,
             \int\,\frac{d^3k\,(1+n(\omega_k))}{2\omega_k(2\pi)^3}\,
       (2\pi)^4\delta^4(q-p-k)
       \sum_{spin}|M(\pi^0\to 2\tilde{\gamma})|^2\\ \nonumber
  & &  \sum_{spin}|M(\pi^0\to 2\tilde{\gamma})|^2=2\,
           A_{\tilde\gamma}^2\,(p\cdot k)^2 \,.
\end{eqnarray}
The factor 1/2 accounts for the two identical bosons in the
final state. The distribution of photons in the final state is Bose
enhanced. The constant $A_{\tilde\gamma}$ is related to the coefficient of the 
electromagnetic anomaly for the isospin axial vector current. 
Anomalous electromagnetic processes in the CFL phase can be 
studied in analogy with the vacuum case~\cite{casal,Nowak:2000wa}  
through the replacement $e\to \tilde{e}=e\,{\rm cos}\,\theta$.
We have
\begin{equation}
  \partial_{\mu}j_5^{\mu\,3}=-\frac{\tilde{e}^2}{32\pi^2}
          \epsilon^{\alpha\beta\mu\nu}\tilde{F}_{\alpha\beta}
          \tilde{F}_{\mu\nu} \,,
\end{equation}
which leads to the identification $A_{\tilde\gamma}=\tilde{e}^2/4\pi^2f_{\pi}$
\cite{peskin}. 
Unlike in vacuum, the matrix element
 $(p.k)^2$ now depends on the pion momentum: 
\begin{equation}
(p.k)^2=\frac{1}{4v_{\pi}^4}
\biggl(m_{\pi}^4 + 
(1-v_{\pi}^2)^2\omega_q^4-2m_{\pi}^2(1-v_{\pi}^2)\omega_q^2\biggr) \,.
\end{equation}
In vacuum, this expression reduces to $m_\pi^4/4$, which is also the
case for a pion with zero momentum in the rest frame of dense matter. 
Note that the
matrix element vanishes if either of the photons become soft. This helps to 
tame potential divergences from the Bose enhancement factors. The
emissivity calculation proceeds similarly to that of the electroweak
decay, where we had ignored the electron mass in the kinematics. Rescaling
energies by the temperature as $x=\omega_q/T, ~y=\omega_p/T,~{\rm and}~ 
\psi=m_{\pi}/T$, we obtain
\begin{equation}
\epsilon_{\tilde\gamma}=\frac{A_{\tilde\gamma}^2T^7}{2^7~\pi^3~v_{\pi}^2}
\int_{\psi}^{\sqrt{\frac{3}{2}}\psi}dx~\frac{x}{{\rm e}^{x}-1}
\biggl(\frac{\psi^2}{v_{\pi}^2} - 
\left(\frac{1-v_{\pi}^2}{v_{\pi}^2}\right)x^2\biggr)^2
\int_{y_{min}}^{y_{\max}} 
dy~\frac{1}{1+e^{-x}-{\rm e}^{-y}-{\rm e}^{y-x}} \,.
\end{equation} 
For the restricted range of $y$ and $x$ above, the exponential factors
in the $y$ integral can be dropped, which is equivalent to neglecting 
Bose enhancement factors. Any possible divergence from this
factor is tamed by vanishing matrix elements and 
phase space. The $y$ integral is then
easily performed yielding 
\begin{equation}
\epsilon_{\tilde\gamma} = 
\frac{A_{\tilde\gamma}^2T^7}{2^7~\pi^3~v_{\pi}^3}
\int_{\psi}^{\sqrt{\frac{3}{2}}\psi}dx~x~e^{-x}
\sqrt{x^2-\psi^2}\biggl(\frac{\psi^2}{v_{\pi}^2} - 
\left(\frac{1-v_{\pi}^2}{v_{\pi}^2}\right)x^2\biggr)^2 \,. 
\end{equation}
With the substitution $x=\psi~{\rm cosh}~\theta$, the above integral can be
expressed as a combination of modified Bessel functions
$K_{\nu}(\psi)$, in the limiting case of 
$m_\pi/T\gg 1$. As their argument is very large, we can utilize the
asymptotic forms of $K_\nu$ to obtain the result
\begin{equation}
\epsilon_{\tilde\gamma}=\frac{A_{\tilde\gamma}^2m_\pi^4}{64\pi}~\frac{1}{v_{\pi}^3}
\biggl(\frac{m_{\pi}T}{2\pi}\biggr)^{3/2} 
 e^{-m_{\pi}/T} 
= \frac{A_{\tilde\gamma}^2m_\pi^4}{64\pi}~n_\pi
\label{eps_gg} \,. 
\end{equation}For purposes of numerical
estimation, we note that for the typical central densities of
(5-10)$n_0$, where $n_0$ is the nuclear saturation density,
the value of the strong coupling $g$ is still much larger than $e$,
so that $\theta_{CFL}$ is small. The emissivity can then be 
expressed as
\begin{equation}
\epsilon_{\tilde\gamma} = 
5.1\times 10^{40}~{m_{10}^4(m_{10}T_9)^{3/2}}{\mu_{100}^{-2}}~
  e^{-\frac{116m_{10}}{T_9}}~{\rm erg~cm^{-3}~s^{-1}} \,.
\end{equation} 
Since thermal pions have an extremely small number density at keV
temperatures, the photon emissivity is negligible despite the large
cross-section for this process.

It is amusing to estimate the mean free path of photons due to
the inverse process $\tilde{\gamma}+\tilde{\gamma}\rightarrow
\pi^0$. 
For the purpose of illustration only, we consider a thermal distribution of 
photons.  
The thermally averaged mean free path may be written as 
\begin{equation}
  \langle\lambda\rangle = v_{\tilde{\gamma}}\langle\tau\rangle \,,
\end{equation}
where the photon velocity is~\cite{Litim}
\begin{equation}
 v_{\tilde{\gamma}} =c/\sqrt{1+\tilde{\kappa}}, \quad
 \tilde{\kappa} = \frac{\tilde{e}^2}{18\pi^2}
         \frac{\mu^2}{\Delta^2} \ll 1\quad.
\end{equation}
Setting $v_{\tilde\gamma}\approx c$ and utilizing the thermal average
of the inverse rate 
\begin{equation}
 \frac{1}{\langle\tau\rangle} = 
    \frac{1}{n_{\tilde\gamma}}\int\,\frac{d^3p\,n(\omega_p)}{(2\pi)^3}
    \frac{1}{\tau(\omega_p)} \label{invtauavg}\quad,
\end{equation}
where $1/\tau(\omega_p)$ is the typical inverse lifetime of a photon
interacting with other thermal photons to form a $\pi^0$ excitation, 
we find
\begin{equation}
\langle\lambda\rangle = 
 \frac{64~\pi c}{A_{\tilde\gamma}^2m_{\pi}^3}~ 
\frac{n_{\tilde\gamma}}{n_{\pi}}\quad
. \label{mfp}
\end{equation}
This result clearly displays the dependences on the vacuum decay rates and 
on the baryon density  
(through $f_\pi$ and $m_{\pi}$ in $A_{\tilde\gamma}$) and temperature 
(through $n_{\tilde\gamma}$ and $n_\pi$). 
Inserting the expression for the number density of thermal photons, 
$n_{\tilde\gamma}=2\xi(3)T^3/\pi^2$, into Eq.~(\ref{invtauavg}), we obtain
\begin{equation}
\langle\lambda\rangle = 1.89\times
10^{-7}~{\mu_{100}^2T_9^{3/2}}{m_{10}^{-9/2}}
 e^{116\frac{m_{10}}{T_9}}\quad {\rm cm} \,.
\end{equation}
For the temperatures of interest, this number is
exponentially large.  Thus, photons in the CFL phase are extremely unlikely to
recombine into a $\pi^0$.

Unlike photon-photon interactions, the processes that contribute
dominantly to the mean free path are Compton
scatterings off charged mesons in the CFL phase.  Since photons stem from the
decay of thermal pions, most of the
photons would have energies well below $m_{GB}$, since
$T\ll m_{GB}<\Delta$.  Thus, the Compton cross-section may be taken
to be adequately represented by the Thomson scattering limit: 
\begin{equation}
\sigma_{\tilde\gamma\pi}\sim\sigma_T^e\biggl(\frac{m_e}{m_{GB}}\biggr)^2\sim
\frac{1}{400}\sigma_T^e\quad;\quad 
\sigma_T^e=\frac{8}{3}\biggl(\frac{e^2}{m_ec^2}\biggr)^2 = 
				66.5~{\rm fm}^{2}\quad.
\end{equation}
The mean free path from scattering off charged pions can be estimated
as
\begin{equation}
l\simeq\frac{1}{\sigma_{\tilde\gamma\pi}~n_{\pi}} 
\sim 2\times 10^{40} ~{\rm km}\,.
\end{equation}
Since the density of scatterers, charged Goldstone bosons, is
exponentially small, the mean free path is very large.  Hence, at low
temperatures, the CFL phase can be regarded as being transparent to
photons \cite{alf} even in the presence of Goldstone bosons.

The mean free path of a photon is essentially the size of the
CFL quark core of the neutron star, since photons equilibriate
rapidly by interacting with the surrounding ordinary matter.
For normal nuclear matter at a density $n_B=3n_0$, with
$n_0=0.16~{\rm fm}^{-3}$, a similar estimate as above yields
\begin{equation}
\l^{\prime} \simeq \frac{1}{\sigma_T^B~n_B}\sim 10^{-8}~{\rm cm}\,.
\end{equation}

We conclude that both photon emissivities and opacities in the CFL
phase are negligible at the temperatures relevant for the long-term
cooling of a neutron star.

\section{Specific Heat of CFL Matter}
\label{sec-cv}

	In order to determine the cooling history of a neutron star, 
we also need the specific heat of CFL matter. The specific heat
of ordinary quark matter is dominated by the quark contribution
\begin{equation}
\label{c_q}
 c_q = \sum_{f=u,d,s} \pi^2n_f\biggl(\frac{k_BT}{\mu_f}\biggr)=
  2.44\times 10^{20} ~T_9 \left(\frac{n_B}{n_0}\right)^{2/3}
   {\rm erg}\,{\rm cm}^{-3}\,{\rm K}^{-1}\,,
\end{equation}
where for simplicity all quarks
have been taken as massless and thus $\mu_f$ is the same for all 
quark flavors. 

The gluon contribution is 
\begin{equation}
\label{c_g}
 c_g = N_g\frac{4\pi^2}{15}T^3
  = 3 N_g \times 10^{13}~T_9^3~ 
 {\rm erg}\,{\rm cm}^{-3}\,{\rm K}^{-1},
\end{equation}
with $N_g=N_c^2-1=8$. The photon contribution is identical to
the gluon contribution with $N_\gamma=1$. 

In ordinary quark matter, there is a non-zero electron chemical
potential. As a result, electrons provide a significant contribution
to the specific heat: 
\begin{equation}
\label{c_e}
 c_e = \pi^2n_e\biggl(\frac{k_BT}{\mu_e}\biggr)=
  0.56\times 10^{20} ~T_9 \left(\frac{Y_e n_B}{n_0}\right)^{2/3}
   {\rm erg}\,{\rm cm}^{-3}\,{\rm K}^{-1}.
\end{equation}

In CFL quark matter, the contributions from quarks 
are suppressed by $\exp(-\Delta/
(k_BT))$, where $\Delta$ is the gap in the appropriate channel. 
In addition, $Y_e \ll 1 $ in the CFL phase. Note, however, that  
at  $T_9=1$, $c_e > c_g+c_\gamma$ till $Y_e=10^{-9}$.

The specific heat of CFL matter receives significant 
contributions from the Goldstone modes. 
The contribution of massive
thermal bosons (that obey the dispersion relation $\omega^2 =
v_\pi^2k^2 + m^2 $) in the CFL phase is modified by the factor
$1/v_\pi^3$ relative to the case with $v_\pi=1$.  For example, for
pions in the non-degenerate limit, we have
\begin{equation}
 c_{GB}(m \neq 0) = \frac 32 n_\pi 
		\left(1 + \frac 52 \frac {T}{m} + \cdots \right)
  = 7.12 \times 10^{15}~(m_{10}T_9)^{3/2} 
     e^{-\frac{116m_{10}}{T_9}}
 ~{\rm erg}\,{\rm cm}^{-3}\,{\rm K}^{-1}\,.
\end{equation}

The contribution of a 
Goldstone mode with dispersion relation $\omega_{GB}=v_\pi k$
is also modified by a factor $1/v_\pi^3$ over that 
of a massless mode with $\omega=k$. With $v_\pi=1/{\sqrt {3}}$, we find
\begin{equation}
 c_{GB} = \frac{6\sqrt{3}\pi^2}{15}T^3
  = 7.8\times 10^{13} ~T_9^3 
 ~{\rm erg}\,{\rm cm}^{-3}\,{\rm K}^{-1}.
\end{equation}
At temperatures on the order of several MeV, there are 
10 almost massless Goldstone modes. When the temperature
drops below 1 MeV, the specific heat of CFL matter is 
dominated by the exactly massless $U(1)_B$ Goldstone
boson. If CFL matter is kaon condensed, then there 
are two massless Goldstone bosons. 

\section{PRINCIPAL FINDINGS}
\label{sec-sum}

 	We have studied neutrino emission from the decay and
scattering processes of Goldstone modes in high density quark matter
in the superconducting color-flavor locked (CFL) phase.  Such a phase
might occur in the core of a neutron star whose mass is near
its maximum allowed value. 

The neutrino emissivities from the new processes that we
have identified scale as
\begin{eqnarray}
\pi^{\pm}\rightarrow l^{\pm} + \bar{\nu_{l}}: \quad
\epsilon_\pi &\sim& (G_F^2\, f_{\pi}^2\, m_e^2)\, 
m_\pi^2\,n_{\pi}\left( 1 + 2\left(\frac{\delta f_\pi}{f_\pi}\right)
      +  \frac{2m_\pi T}{v_{\pi}^2m_e^2}\left(\frac{\delta f_\pi}{f_\pi}\right)^2
   \right)\, \nonumber   \\
K^{\pm}\rightarrow l^{\pm} + \bar{\nu_{l}}: \quad
\epsilon_{K^\pm} &\sim& (\sin^2\theta_C~G_F^2\, f_{\pi}^2\, m_e^2)\, 
E_{K^\pm}^2\,n_{K^\pm}\nonumber \\
&\times&\left( 1 + 2\left(\frac{\delta f_\pi}{f_\pi}\pm\frac{\mu_{eff}}{E_{K^\pm}}\right)
      +  \frac{2m_{K^\pm} T}{v_K^2m_e^2}\left(\frac{\delta f_\pi}{f_\pi}\pm\frac{\mu_{eff}}{E_{K^\pm}}\right)^2
   \right)\,  \nonumber  \\
\pi^0(\eta,\eta')\rightarrow \nu + \bar{\nu}: \quad
\epsilon_{\nu\bar{\nu}} &\sim& (G_F^2m_\pi^2\delta f_{\pi}^2)\,T\,m_{\pi}
\,n_{\pi} \, \nonumber  \\
\phi+\phi\to \phi +\nu+\bar{\nu}: \quad
 \epsilon_{GB} &\sim&   \frac{G_F^2}{f^2\mu^4} T^{15} \,,
\end{eqnarray} 
where $\delta f_\pi = f_T-f_S$ with $f_S=f_\pi/3$,~and $n_\pi$ and $n_K$
are the number densities of pions and kaons, respectively.  Since the
masses of these Goldstone bosons are expected to be of order 10 MeV
and $T \leq 0.1$ MeV, these Goldstone bosons are 
non-degenerate and obey Boltzmann statistics. Hence, their
number densities scale as $(mT)^{3/2} \exp(-m/T)$.  This is in
contrast to quarks, which are degenerate  since
$\mu_B/T \gg 1$. However, emissivities from gapped quarks are suppressed by
a factor $\exp(-\Delta/T)$, with the gap $\Delta$ being of order 100
MeV in the CFL phase.  The hierarchy of physical scales, $\mu_B > 
\Delta > m_{GB} > T$ (for the most part, we have taken $\mu_e = 0$ in
the CFL phase) thus assures that the neutrino emissivities from
massive Goldstone modes dominate over those from gapped quarks.

The unique feature of the $\pi^0\to\nu+\bar{\nu}$ process is that it
requires Lorentz symmetry breaking, which manifests itself in the
difference of the spatial and temporal pion decay constants, $f_S$ and
$f_T$.  This emissivity is thus proportional to $\delta f_{\pi}^2$.
In vacuum, helicity conservation forbids such a decay, but in matter
where boost invariance is lost, this process is allowed for all non-zero 
pion momenta. 

The emissivities from the massive Goldstone modes are expressible in
terms of simple products $~{\rm interaction~ strength} \times ~
{\rm relevant~ energies} ~ \times ~{\rm the~ number~ density~ of~
thermal~ bosons}$, mainly because bosons are non-degenerate.  Such
simplicity is lost when these modes exist in partially degenerate or
degenerate conditions because of Bose enhancement effects.  It must be
stressed that the emissivities from the massive Goldstone modes, even
though they dominate those from gapped quarks, remain rather small,
chiefly because of their very small number densities at the
temperatures of relevance to long-term cooling. For example, at
$T_9=1$, $n_\pi \sim 3.5 \times 10^{-58}~{\rm fm}^{-3}$.

Neutrino emission from the scattering of massless Goldstone modes
(numbering one in the CFL phase and two in the kaon condensed phase)
is not exponentially penalized. However, this process appears only
at ${\cal O}(p^4)$ in a chiral effective Lagrangian, which leads to
a numerically small emissivity.  Nevertheless, it is likely to be the
principal source of neutrino emission from the CFL phase.

We have also calculated the photon emissivity from the CFL phase. The
anomalous decay process $\pi^0\to{\tilde\gamma} +\tilde{\gamma}$
yields an intuitive expression for the emissivity,
$\epsilon_{\tilde\gamma}=({A_{\tilde\gamma}^2m_\pi^4}/{64\pi})~n_\pi$,
when $m_{\pi}\gg T$. This emissivity may be readily understood as the
product of the energy per unit time carried by the photons times the
number density of thermal pions from which they are
emitted. Numerically, the emissivity is exponentially small, not
because the cross-section for the anomalous process is small, but
because, at $T_9=1$, the thermal pion density is exceedingly small.
Our estimation of the photon mean free path, due to Compton scattering
in the Thomson limit, shows that the CFL phase is transparent to
photons even in the presence of Goldstone bosons.  However, photons
are thermalized quickly in the high density normal baryonic matter
that surrounds the CFL phase. 

Since electrons are strictly absent in bulk CFL matter, its specific
heat per unit volume $c_V$ is dominated by the massless $U(1)$ mode
associated with superfluidity (two such modes exist for the kaon
condensed phase). This bosonic excitation ($\phi$) is relatively
easier to excite than all other massive Goldstone modes, which are
exponentially suppressed by thermal occupation factors. Thermal
gradients arising in the CFL phase are rapidly reduced due to the
large thermal conductivity, which is a consequence of small
cross-sections for $\phi\phi$ scattering as well as their low number
density. As expected, the transport properties of the CFL phase are
controlled essentially by its superfluid nature.

\section{IMPLICATIONS FOR THE COOLING OF COMPACT STARS}
\label{sec-cooling}
                                                                                                                                                                                        
 	Our results for the neutrino emissivity in the CFL phase lead
naturally to the issue of their efficiency in cooling the {\it
{interior}} of the star.  The thermal evolution is determined by the 
equations of radiative transport and energy balance which may be cast to 
read as~\cite{Lattimer94}  
\begin{equation}
\frac {1}{r^2}\frac {\partial}{\partial r}
\left (r^2 K \frac{\partial T}{\partial r} \right) 
= c_V  \frac{\partial T}{\partial t} = -(\epsilon + H) 
\,,
\label{transport}
\end{equation}
where $T$ is the temperature, $r$ is the radial coordinate (general
relativistic corrections have been suppressed for simplicity), $K$ is
the thermal conductivity, and $c_V$ is the total specific heat per
unit volume. The total emissivity due to photons and neutrinos is
denoted by $\epsilon$ and possible internal heating sources are
contained in $H$.  We can estimate the cooling timescales $\Delta t$
of the CFL phase in isolation with the expression
\begin{equation}
\Delta t = - \int_{T_i}^{T_f} dT~\frac {c_V}{\epsilon} \,.
\end{equation}
In a strictly electron-free CFL phase, the largest contributions to
the specific heat comes from those of gluons, photons, and the
massless Goldstone modes, since those of the massive Goldstone modes
are severely suppressed.  For the same reason, the emissivity is
dominated by the scattering of massless Goldstone modes.  Using $c_V
\simeq 3.5 \times 10^{14}~T_9 ~{\rm erg}\,{\rm cm}^{-3}\,{\rm K}^{-1}$
and $\epsilon \simeq 10^{-11} \times T_9^{15} ~{\rm
erg~cm^{-3}~s^{-1}}$ from Secs. II and III, we find that $\Delta t
\sim 10^{26}~T_9^{-11}\mu_{100}^6$ y.  This time scale
is extremely long!

Note that, were the massive Goldstone bosons to operate in isolation, 
the cooling time would be vastly different. Using $c_V \simeq 7.2 \times
10^{15}~ (m_{10}T_9)^{3/2}~\exp(-116m_{10}/T_9)~ ~{\rm erg}\,{\rm
cm}^{-3}\,{\rm K}^{-1}$ and $\epsilon \simeq 1.2 \times T_9^{28}
\mu_{100}^2 m_{10}^2 (m_{10}T_9)^{3/2}~\exp(-116m_{10}/T_9)~ ~{\rm
erg~cm^{-3}~s^{-1}}$, we obtain $\Delta t = 6\times 10^{-4} \times 
(T_9(i) - T_9(f)) \\ \mu_{100}^{-2}~ m_{10}^{-2}~{\rm s}$.  In this case,
$c_v/\epsilon$ becomes independent of $T$, which leads to a 
cooling rate that is linear in $T$, and also exceedingly rapid.

The CFL phase is characterized by a very large thermal conductivity
due to the smallness of both the Goldstone boson number densities and
cross sections in CFL matter. The temperature gradient in the CFL core
will thus be very small, and the CFL core will be nearly isothermal,
with its temperature pinned to that of the surrounding hadronic matter
(this is easily verified by seeking separable solutions to
Eq.~(\ref{transport})). In the realistic case in which a neutron star
has a small CFL core, the thermal content and the emissivity of the
CFL core will be negligible compared to those of the surrounding
hadronic matter.\footnote{In baryonic matter, the simplest possible
$\nu$ emitting processes are the direct Urca processes $f_1 + \ell
\rightarrow f_2 + \nu_\ell\,, f_2 \rightarrow f_1 + \ell +
\overline{\nu_\ell}$, where $f_1$ and $f_2$ are baryons and $\ell$ is
either an electron or a muon.  These processes can occur whenever
momentum conservation is satisfied among $f_1, f_2$ and $\ell$.  If
the unsuppressed direct Urca process for {\em any} component occurs, a
neutron star will rapidly cool because of large energy losses due to
neutrino emission: the star's interior temperature $T$ will drop below
10$^9$ K in minutes and reach 10$^7$ K in about a hundred years.  This
is the so-called rapid cooling paradigm~\cite{Lattimer91,Prakash92}.
If no direct Urca processes are allowed, or they are all suppressed
due to baryon superfluidity (with gaps of order 1 MeV), cooling
instead proceeds through the significantly less rapid modified Urca
process in which an additional fermion enables momentum conservation.
This situation could occur if no hyperons are present, or the nuclear
symmetry energy has a weak density
dependence~\cite{Lattimer91,Prakash92}.  Comparatively less rapid, but 
faster than that due to modified Urca cooling occurs in the
presence of Bose condensates~\cite{cool}.}  The picture that then
emerges is that such a star will cool nearly identically to a star
without a CFL core. The presence of a CFL phase in dense matter will
apparently leave no observable trace in the long-term cooling history
of the star.

The interface between electron-rich nuclear matter and
electron-deficient CFL quark matter was analyzed in
Ref.~\cite{Alford:2001zr} with the conclusion that a thin charged
layer ($\sim 10$ fm) separates the two bulk phases if the surface
energy $\sigma_s$ is large (dimensional estimates place $\sigma_s\sim
400$ MeV, which is too costly for a mixed phase). In this case,
electrons can leak in to the CFL phase.  Owing to the partially
inverted mass spectrum of Goldstone bosons in the CFL phase (in
particular, $m_K<m_\pi$), electrons can decay into kaons through
electroweak interactions near the CFL side of the interface, resulting
in neutrino emission and the formation of a $K^-$ condensate.  This
causes the proton concentration on the nuclear side of the interface
to rise substantially in order to maintain charge
neutrality. Consequently, there is the possibility that, within this
very thin layer, the direct Urca process $n \rightarrow p+e^-+\bar
\nu$ would occur as long as the proton fraction is of order $11-14\%$
for which momentum conservation between the participating fermions
becomes possible~\cite{Lattimer91}.  Inasmuch as this thin interface
is embedded within the star, the occurence of a such a rapid cooling
process would be masked by cooling from the much thicker exterior
layers in which similar processes are also possible. From an
observational standpoint, such cooling would resemble cooling from
normal stars.

In Ref.~\cite{Alford:2001zr}, the phase structure of baryon and CFL
quark matter was also studied as a function of the magnitude of
$\sigma_s$.  For a sufficiently small $\sigma_s$, a homogeneous mixed
phase with domains of CFL quark matter and nuclear matter was found to
be preferred over a wide range of density.  In this case, neutrino
emission would be dominated by the nuclear bubbles which can cool via
the rapid processes characteristic of the nuclear phase (see footnote
3).

Recently, the cooling of a self-bound quark star with a bare surface
has been studied in Ref.~\cite{Page02} by considering both the CFL and
2SC phases. If the entire star is in the CFL phase, which permits no
electrons, cooling occurs chiefly through processes that involve the
Goldstone modes. As noted above, the cooling times for this case are
too long to be observable.  If the CFL phase is surrounded by the 2SC
phase at lower densities, cooling is dominated by neutrino emission
from direct Urca processes involving unpaired quarks in the 2SC phase
(which allows electrons to exist) and by the annihilation of $e^+e^-$
pairs produced by an intense electrical field which binds electrons to
the surface of quark matter. This is essentially the Schwinger
mechanism in the presence of a Fermi sea of electrons, and can occur
only at finite temperature~\cite{VVU98}. Beginning at a temperature of
$\sim 10^{11}$K ($\sim 10$ MeV) down to $\sim 10^8$K, the temperature
versus age curve for the case with the 2SC phase in which gaps are of
order 100 MeV differs little from that of the normal phase, chiefly
because pairing reduces neutrino emiision and the specific heat by
similar amounts.  Substantial differences from the cooling of the
normal phase occur, however, if quarks not participating in 2SC
pairing also pair through residual interactions, but with gaps of
order 1 MeV. In this case, the hard X-ray spectrum with a mean energy
of $\sim 10^2$ keV is proffered as an observational signature of bare
quark stars.

In this work, we have focused on the long-term cooling epoch when most
neutrinos have left the star and temperatures are in the range of
hundreds of keV to tens of eV. In a proto-neutron star, neutrinos are
trapped and temperatures range from 20-40 MeV. The CFL phase is likely
to appear during the PNS evolution.  During this stage, $m_{GB}/T\le
1$ and hence the Goldstone modes are in the partially degenerate or
degenerate regime.  As $m_{GB}/T\ \ll \Delta/T$, both neutral and
charged current neutrino interactions with the Goldstone bosons offer
new sources of opacity.  Some examples of possible processes are
$\nu(\bar{\nu})+\pi^{\pm}\to \nu(\bar{\nu}) +\pi^{\pm}$, $\nu + \pi^-
\to e^- + \pi^0$, $\bar{\nu} + \pi^+ \to e^+ + \pi^0$ and
$\nu(\bar{\nu}) + \pi^0 \to e^{\mp} + \pi^{\pm}$.  Note that with fast
neutrinos, the processes $\nu \to \pi^0 + \nu$ and $\bar{\nu} \to \pi^+ +
e^-$ also become allowed and contribute to the opacity.  The resultant
neutrino opacities are expected to be significantly larger than those
from scattering and absorption on gapped quarks, since temperatures
are high enough to admit large pion densities.  For example, in the
range $20 < T/{\rm MeV} < 40$, the number densities of a given species
of a Goldstone mode whose mass lies in the range $0 < m_{GB}/{\rm MeV}
< 10$ are $n_{GB}\sim 10^{-3}-10^{-4}~{\rm fm}^{-3}$, substantially
more than those encountered in the long-term cooling epoch.

\vspace*{0.1in}

\noindent {\bf NOTES:} (a) Concurrently and independently of our work,
Reddy, Sadzikowsky, and Tachibana~\cite{Reddy02} have performed a
detailed study of $\nu$-opacities and emissivities in the CFL phase at
tens of MeV temperatures encountered in the evolution of a
proto-neutron star.  Their quantitative results underscore the
important role of Goldstone boson excitations in the CFL phase and
provide the groundwork for calculations of observable neutrino
luminosities to be performed.  (b) Shovkovy and
Ellis~\cite{Shovkovy02} have performed a calculation of the thermal
conductivity in the CFL phase at temperatures of relevance to
long-term cooling.  They find that the dominant contribution to the
conductivity comes from photons and Goldstone bosons.  This result is
a necessary ingredient in quantitative calculations of a compact
star's long-term cooling.  (c) Page and Usov~\cite{Page02} have
calculated the thermal evolution and light curves of young bare
(self-bound) quark stars considering both the 2SC and CFL phases. They
find that the energy gap of superconducting quark matter may be
estimated from the light curves if it is in the range  $\sim 0.5$ MeV
to a few MeV.

\vspace*{0.1in} 
%Acknowledgements: 
We would like to thank P.~Bedaque and J.M.~Lattimer for useful
discussions.  We are grateful to K.~Rajagopal for his comment
concerning the thermal conductivity of the CFL core and to S.~Reddy
for helpful remarks with regard to emissivities involving kaons.  This
work was supported in part by the US DOE grant DE-FG-88ER40388.

%-----------------------THE REFERENCES------------------------------------
\begin{flushleft}

\end{flushleft}
\end{document}